\documentclass[a4paper,UKenglish]{lipics-v2021}
\usepackage{microtype,mathdefs,array,amsthm,enum,relsize,booktabs}
\usepackage[section]{thmdefs}

\nolinenumbers

\makeatletter
\newcommand\m@thsm@ller[2]{\mbox{\relscale{0.91}$\m@th#1#2$}}
\let\smaller\undefined
\DeclareRobustCommand\smaller[1]{\relax\ifmmode{\mathpalette\m@thsm@ller{#1}}\else{\relscale{0.91}#1}\fi}
\makeatother

\DeclareRobustCommand*{\dom}{\qopname\relax o{dom}}

\newcommand*{\id}{\mathrm{id}}

\newcommand*{\ar}{\mathrm{ar}}

\newcommand*{\Flat}{\mathrm{flat}}
\newcommand*{\sing}{\mathrm{sing}}

\newcommand*{\CTL}{\smaller{\mathrm{CTL}}}

\newcommand*{\EF}{\smaller{\mathrm{EF}}}
\newcommand*{\cEF}{\mathrm{c\smaller{\mathrm{EF}}}}

\DeclareRobustCommand*{\tp}{\qopname\relax o{tp}}
\DeclareRobustCommand*{\Tp}{\qopname\relax o{Tp}}

\newcommand*{\nasymp}{\not\asymp}

\newcommand*{\?}{\kern .08em}

\newcommand\upqed{\vskip-\baselineskip\vskip-\belowdisplayskip}
\newcommand\markenddef{\hfill$\lrcorner$}

\title{$\omega$-Forest Algebras and Temporal Logics}
\titlerunning{\boldmath$\omega$-Forest Algebras and Temporal Logics}
\author{Achim Blumensath}{Masaryk University Brno}{blumens@fi.muni.cz}{}{Work supported by the Czech Science Foundation, grant No.~GA17-01035S}
\author{Jakub Lédl}{Masaryk University Brno}{jakubledl@mail.muni.cz}{}{Work supported by the Czech Science Foundation, grant No.~GA17-01035S}
\authorrunning{A. Blumensath,  J. Lédl}
\begin{document}
\maketitle

\Copyright{Achim Blumensath, Jakub Lédl}

\begin{abstract}
We use the algebraic framework for languages of infinite trees introduced in~\cite{Blumensath20}
to derive effective characterisations of various temporal logics, in particular
the logic $\EF$ (a~fragment of $\CTL$) and its counting variant $\cEF$.
\end{abstract}

\section{Introduction}   

Among the many different approaches to language theory, the algebraic one
seems to be particularly convenient when studying questions of expressive power.
While algebraic language theories for word languages (both finite and infinite)
were already fully developed a long time ago,
the corresponding picture for languages of trees, in particular infinite ones,
is much less complete.
Seminal results contributing to such an algebraic framework for languages of infinite trees
were provided by the group of Boja\'nczyk \cite{BojanczykId09,BojanczykIdSk13}
with one article considering languages of \emph{regular} trees only,
and one considering languages of \emph{thin} trees.
The first complete framework that could deal with arbitrary infinite trees
was provided in~\cite{Blumensath11c,Blumensath13a}.
Unfortunately, it turned out to be too complicated and technical for applications.
Recently, two new general frameworks have been introduced \cite{BlumensathZZ,Blumensath20}
which seem to be more satisfactory\?:
one is based on the notion of a \emph{branch-continuous tree algebra,} while the other
uses \emph{regular tree algebras.}
The first one seems to be more satisfactory from a theoretical point of view,
while the second one is more useful for applications, in particular for characterisation results.

In this article we concentrate on the approach based on regular tree algebras
from~\cite{Blumensath20} which seems to be emerging as the standard.
The goal is to apply the framework to a few test cases and to see how well it
performs for its intended purpose.
While the definition of a regular tree algebra (given in Section~\ref{Sect: forest algebras} below)
is a bit na\"ive and seems circular at first sight,
it turns out that it is sufficient to guarantee the properties we need for applications\?:
one can show that (i)~the class of regular tree algebras forms a pseudo-variety and
that (ii)~every regular tree language has a syntactic algebra, which is in fact a regular
tree algebra.
By general category-theoretic results, such as those from
\cite{Bojanczyk15,Bojanczyk20}~or~\cite{Blumensath21},
this implies that there exists a Reiterman type theorem for such algebras, i.e.,
the existence of equational characterisations for sub-pseudo-varieties.
This is precisely what is needed for a characterisation theorem.

Unfortunately progress on an algebraic theory of infinite trees has been rather slow
since matters have turned out to be significantly
more complicated than the case of words or finite trees.
Hence, every step of progress is very welcome.
For instance, the recent paper~\cite{ColcombetJaquard21} characterises the languages of infinite
trees that are recognised by algebras of bounded growth.
The applications we are looking at in the present paper concern certain temporal logics,
in particular, the logic $\EF$ and its counting variant $\cEF$,
and we aim to derive decidable algebraic characterisations for them using our algebraic
framework.
Note that Boja\'nczyk and Idziaszek have already provided a decidable characterisation
for $\EF$ in~\cite{BojanczykId09}, but their result is only partially algebraic.
They prove that a regular language is definable in~$\EF$ if, and only if,
the language is bisimulation-invariant and its syntactic algebra satisfies a certain equation,
but they were not able to provide an algebraic characterisation of bisimulation invariance.
Due to our more general algebraic framework we are able to fill this gap below.

We start in the next section with a short overview of the algebraic framework
from~\cite{Blumensath20}.
We have to slightly modify this material since it was originally formulated in the setting
of ranked trees while, when looking at temporal logics, it is more natural to consider
unranked trees and forests.
The remainder of the article contains our various characterisation results.
In Section~\ref{Sect: bisimulation} we derive an algebraic characterisation of
bisimulation-invariance, the result missing in~\cite{BojanczykId09}.
Then we turn to our main result and present characterisations for the logic $\cEF$
and some of its fragments, including the logic $\EF$.
These results and some of their consequences are presented in
Section~\ref{Sect: cEF}, while the proofs are deferred to Section~\ref{Sect: cEF proof}.

\section{Forest algebras}   
\label{Sect: forest algebras}

The main topic of this article are languages of (possibly infinite) forests
and the logics defining them. Before introducing the algebras we will use to recognise such
languages, let us start by fixing some notation and conventions.
Although our main interest is in unranked forests, we will use a more general version
that combines the ranked and the unranked cases. As we will see below
(cf.~Theorem~\ref{Thm: bisimulation invariance}), the ability to use ranks will
increase the expressive power of equations for our algebras considerably.
Thus, we will work with \emph{ranked sets,} i.e., sets where every element~$a$ is assigned
an \emph{arity} $\ar(a)$. Formally, we consider such sets as families $A = (A_m)_{m<\omega}$,
where $A_m$~is the set of all elements of~$A$ of arity~$m$.
Functions between ranked sets then take the form $f = (f_m)_{m<\omega}$ with $f_m : A_m \to B_m$.

We will consider (unranked, finitely branching, possibly infinite) forests where each vertex
is labelled by an element of a given ranked set~$A$ and each edge is labelled by a
natural number with the restriction that, if a vertex is labelled by an element of arity~$m$,
the numbers labelling the outgoing edges must be less than~$m$.
If an edge $u \to v$ is labelled by the number~$k$, we will call~$v$ a \emph{$k$-successor} of~$u$.
Note that a vertex may have several $k$-successors, or none at all.
We assume that the roots of a forest are ordered from left to right,
as are all the $k$-successors of a given vertex~$v$,
while we impose no ordering between a $k$-successor and an $l$-successor, for $k \neq l$.
We write $\bbF_0A$ for the set of all such $A$-labelled forests.
(We shall explain the index~$0$ further below.)
We write $\dom(s)$ for the set of vertices of a forest $s \in \bbF_0A$,
and we will usually identify $s$~with the function $s : \dom(s) \to A$
that maps vertices to their labels.
We denote the empty forest by~$0$ and the disjoint union of two forests $s$~and~$t$ by $s+t$
(where the roots of~$t$ are added after those of~$s$).
We will frequently use term notation to denote forests such as
\begin{align*}
  a(b + c, 0, b) + b\,,
\end{align*}
which denotes a forest with two components\?: the first one consisting of a root labelled
by an element~$a$ of arity~$3$ which has two $0$-successors labelled $b$~and~$c$,
no $1$-successor, and one $2$-successor\?; the second component consists of a singleton
with label~$b$.

We use the symbol~$\preceq$ for the forest ordering where the roots are the minimal elements
and the leaves the maximal ones.
For a forest~$s$, we denote by~$s|_v$ the subtree of~$s$ attached to the vertex~$v$.
The \emph{successor forest} of~$v$ in~$s$ is the forest obtained from~$s|_v$ by removing the root~$v$.

For a natural number~$n$, set $[n] := \{0,\dots,n-1\}$.
An \emph{alphabet} is a finite (unranked) set~$\Sigma$ of symbols.
If we use an alphabet in a situation such as $\bbF_0\Sigma$ where a ranked set is expected,
we will consider each symbol in~$\Sigma$ as having arity~$1$.
Thus, for us a \emph{forest language over an alphabet~$\Sigma$} will be a set
$L \subseteq \bbF_0\Sigma$ consisting of the usual unranked forests.
(The power to have elements of various arities is useful when writing down algebraic equations,
but it is rather unnatural when considering languages defined by temporal logics.)
We denote by~$\Sigma^*$ the set of all finite words over~$\Sigma$,
by~$\Sigma^\omega$ the set of infinite words, and $\Sigma^\infty := \Sigma^* \cup \Sigma^\omega$.
A~\emph{family of (word, forest,\ldots) languages} is a function~$\calK$ mapping
each alphabet~$\Sigma$ to a class $\calK[\Sigma]$ of (word, forest,\ldots) languages over~$\Sigma$.

\begin{figure}
\centering
\includegraphics{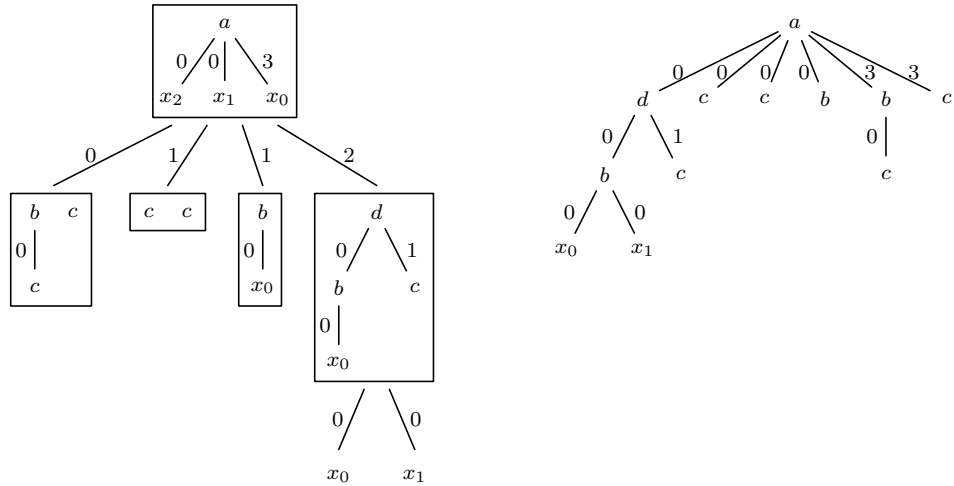}
\caption{The flattening operation}\label{fig: flat}
\end{figure}

Our algebraic framework to study forest languages is built on the notion of an
Eilenberg--Moore algebra for a monad. To keep category-theoretical prerequisites
at a minimum we will give an elementary, self-contained definition.
The basic idea is that, in the same way we can view the product of a semigroup as an
operation turning a sequence of semigroup elements into a single element,
we view the product of a forest algebra as an operation turning a given forest that is
labelled with elements of the algebra into a single element.
The material in this section is taken from~\cite{Blumensath20} with minor adaptations to
accommodate the fact that we are dealing with unranked forests instead of ranked trees.
Proofs can also be found in~\cite{Blumensath21}, although in a much more general setting.
We start by defining which forests we allow in this process.
\begin{Def}
(a)
We denote by $\bbF$ the functor mapping a ranked set~$A$ to the ranked set
$\bbF A = (\bbF_m A)_m$ where $\bbF_m A$ consists of all $(A \cup \{x_0,\dots,x_{m-1}\})$-labelled
forests such that
\begin{itemize}
\item the new labels $x_0,\dots,x_{m-1}$ have arity~$0$,
\item each label~$x_i$ appears at least once, but
  only finitely many times, and
\item no root is labelled by an~$x_i$.
\end{itemize}

(b) The \emph{singleton function} $\sing : A \to \bbF A$ maps a label~$a$ of arity~$m$
to the forest $a(x_0,\dots,x_{m-1})$.

(c) The \emph{flattening function} $\Flat : \bbF\bbF A \to \bbF A$ takes a forest $s \in \bbF\bbF A$
and maps it to the forest $\Flat(s)$ obtained by assembling all forests $s(v)$, for $v \in \dom(s)$,
into a single large forest. This is done as follows. For every vertex of~$s(v)$ that is labelled
by a variable~$x_k$, we take the disjoint union of all forests labelling the $k$-successors of~$v$
and substitute them for~$x_k$. This is done simultaneously for all $v \in \dom(s)$ and all
variables in~$s(v)$ (see Figure~\ref{fig: flat} for an example.)
\markenddef
\end{Def}

Now we can define a forest algebra to be a set~$A$ equipped with a product $\bbF A \to A$.
\begin{Def}
(a) An \emph{$\omega$-forest algebra} $\frakA = \langle A,\pi\rangle$ consists of a ranked
set~$A$ and a function $\pi : \bbF A \to A$ satisfying the following two axioms\?:
\begin{align*}
  \text{the \emph{associative law}}\quad
  \pi \circ \bbF\pi = \pi \circ \Flat
  \quad\qtextq{and}\quad
  \text{the \emph{unit law}}\quad
  \pi \circ \sing = \id\,.
\end{align*}
We will denote forest algebras by fraktur letters~$\frakA$ and their universes
by the corresponding roman letter~$A$. We will usually use the letter~$\pi$ for the product,
even if several algebras are involved.

(b) A \emph{morphism} of $\omega$-forest algebras is a function $\varphi : \frakA \to \frakB$
that commutes with the products in the sense that
$\pi \circ \bbF\varphi = \varphi \circ \pi$.
\markenddef
\end{Def}

\begin{Rem}
(a)
In the following we will simplify terminology by dropping the~$\omega$ and simply speaking
of \emph{forest algebras.} But note that, strictly speaking, this name
belongs to the kind of algebras introduced by Boja\'nczyk and Walukiewicz
in~\cite{BojanczykWalukiewicz07}.

(b) One can show that the functor~$\bbF$ together with the two natural transformations
$\Flat$ and $\sing$ forms what is called a \emph{monad} in category theory.
In this terminology, we can define forest algebras as \emph{Eilenberg-Moore algebras}
for this monad.

(c)
Note that a forest algebra $\frakA = \langle A,\pi\rangle$ contains a monoid
$\langle A_0,{+},0\rangle$ (called the \emph{horizontal monoid}) and an $\omega$-semigroup
$\langle A_1,A_0,{}\cdot{}\rangle$ (the \emph{vertical $\omega$-semigroup}),
whose operations are derived from the product~$\pi$.
For instance, the vertical product $a\cdot b$, for $a,b \in A_1$, is formed as the
produce $\pi(s)$, where $s$~consists of a root labelled~$a$, an internal vertex labelled~$b$,
and a leaf labelled be the variable~$x_0$.

(d) The reason why we do not allow forests where some root is labelled by a variable~$x_k$
is that an infinite product of such forests is not always defined. For instance, multiplying
an infinite sequence of forests of the form $x_0 + a$ would create a forest
with infinitely many components, which is not allowed.
\end{Rem}

Sets of the form $\bbF A$ can be equipped with a canonical forest algebra structure
by using the flattening operation $\Flat : \bbF \bbF A \to \bbF A$ for the product.
By general category-theoretical considerations it follows that algebras of this form are
exactly the \emph{free} forest algebras (generated by~$A$).
In this article we consider \emph{forest languages} over an alphabet~$\Sigma$
as subsets $L \subseteq \bbF_0\Sigma$. Such a language is \emph{recognised} by a morphism
$\eta : \bbF\Sigma \to \frakA$ of forest algebras if $L = \eta^{-1}[P]$ for some $P \subseteq A_0$.
\begin{Exam}
Let $\Sigma := \{a,b\}$.
We can recognise the language $L \subseteq \bbF_0\Sigma$ of all forests~$s$
containing at least one occurrence of the letter~$a$ as follows.
Let $\frakA$~be the algebra consisting of two elements $0_m$~and~$1_m$, for each arity~$m$,
where the product~$\pi$ maps a forest $s \in \bbF_m A$ to~$1_m$ if at least one vertex
is labelled by~$1_n$, for some~$n$. Otherwise, $s$~is mapped to~$0_m$.
Then $L = \varphi^{-1}(1_0)$ where the morphism $\varphi : \bbF_0\Sigma \to \frakA$ is defined
by $\varphi(a) := 1_1$ and $\varphi(b) := 0_1$.
(As~$\bbF\Sigma$ is freely generated by the set $\{a,b\}$,
this determines~$\varphi$ for all inputs.)
\end{Exam}

In analogy to the situation with word languages we would like to have a theorem
stating that a forest language is regular if, and only if, it is recognised by a morphism
into some finite forest algebra. But this statement is wrong for two reasons.
The first one is that every forest algebra with at least one element of positive arity
has elements of every arity and, thus, is infinite. (For instance, given $a \in A_1$,
we obtain an element $a(x_0 +\dots+ x_{n-1}) \in A_n$ of every arity~$n$).
To fix this, we have to replace the property of being finite by that of having only
finitely many elements of each arity. We call such algebras \emph{finitary.}

But even if we modify the statement in this way it still fails since one can find
finitary forest algebras recognising non-regular languages.
(An example for tree languages is given by Boja\'nczyk and Klin in~\cite{BojanczykKlin18}.)
Therefore we have to restrict our class of algebras.
A~simple way to do so is given by the class of (locally) regular algebras
introduced in~\cite{Blumensath20} where all of the following results
are taken from (again in the case of trees instead of forests).
\begin{Def}
Let $\frakA$~be a forest algebra.

(a) A subset $C \subseteq A$ is \emph{regularly embedded} if, for every element $a \in A$,
the preimage $\pi^{-1}(a) \cap \bbF C$ is a regular (i.e., automaton recognisable)
language over~$C$.

(b) $\frakA$~is \emph{locally regular} if every finite subset is regularly embedded.

(c) $\frakA$~is \emph{regular} if it is finitary, finitely generated, and locally regular.
\markenddef
\end{Def}
This definition of a regular forest algebra is not very enlightening.
We refer the interested reader to~\cite{Blumensath20}
for a purely algebraic (but much more complicated) characterisation.
\begin{Thm}
Let $L \subseteq \bbF_0\Sigma$ be a forest language. The following statements are equivalent.
\begin{enum1}
\item $L$~is regular (i.e., automaton recognisable).
\item $L$~is recognised by a morphism into a locally regular forest algebra.
\item $L$~is recognised by a morphism into a regular forest algebra.
\end{enum1}
\end{Thm}
(The reason why we introduce two classes is that locally regular algebras enjoy better
closure properties, while the regular ones are more natural as recognisers of languages.)
One can show (see~\cite{Blumensath20}) that the (locally) regular algebras form a
pseudo-variety in the sense that locally regular algebras are closed under quotients, subalgebras,
finite products, and directed colimits, while regular algebras are closed under quotients,
finitely generated subalgebras, finitely generated subalgebras of finite products, and
so-called `rank-limits'.
More important for our current purposes is the existence of syntactic algebras
and the fact that these are always regular.
\begin{Def}
Let $L \subseteq \bbF\Sigma$ be a forest language.

(a) The \emph{syntactic congruence} of~$L$ is the relation
\begin{align*}
  s \sim_L t \quad\defiff\quad p[s] \in L \Leftrightarrow p[t] \in L\,,
  \quad\text{for every context } p\,,
\end{align*}
where a \emph{context} is a $(\Sigma \cup \{\Box\})$-labelled forest
(where $\Box$~is a new symbol of the same arity as $s$~and~$t$)
and $p[s]$~is the forest obtained from~$p$ by replacing each vertex labelled by~$\Box$
by the forest~$s$.

(b) The \emph{syntactic algebra} of~$L$ is the quotient
$\frakS(L) := \bbF\Sigma/{\sim_L}$.
\markenddef
\end{Def}
\begin{Thm}
The syntactic algebra $\frakS(L)$ of a regular forest language~$L$ exists, it is regular,
and it is the smallest forest algebra recognising~$L$.
Furthermore, $\frakS(L)$ can be computed given an automaton for~$L$.
\end{Thm}

Regarding the last statement of this theorem, we should explain what we mean by
computing a forest algebra.
Since forest algebras have infinitely many elements, we cannot simply compute the full multiplication
table. Instead, we say that a regular forest algebra~$\frakA$ is computable if,
given a number $n < \omega$,
we can compute a list $\langle\calA_a\rangle_{a \in A_n}$ of automata such that
$\calA_a$~recognises the set $\pi^{-1}(a) \cap \bbF C$, for some fixed set~$C$ of generators.

\section{Bisimulation}   
\label{Sect: bisimulation}

To illustrate the use of syntactic algebras let us start with a simple warm-up exercise\?:
we derive an algebraic characterisation of bisimulation invariance.
This example also explains why algebras with elements of higher arities are needed
(this is the reason Boja\'nczyk and Idziaszek~\cite{BojanczykId09}, whose framework
supported only arity~$1$, had to leave a similar characterisation as an open problem).

Recall that a \emph{bisimulation} between two forests $s$~and~$t$ is a binary relation
$Z \subseteq \dom(s) \times \dom(t)$ such that $\langle u,v\rangle \in Z$ implies that
\begin{itemize}
\item $s(u) = t(v)$ and,
\item for every $k$-successor~$u'$ of~$u$, there is some $k$-successor~$v'$ of~$v$ with
  $\langle u',v'\rangle \in Z$ and vice versa.
\end{itemize}
Two trees are \emph{bisimilar} if there exists a bisimulation between them that relates
their roots. More generally, two forests are bisimilar if every component of one is bisimilar
to some component of the other.
A language~$L$ of forests is \emph{bisimulation-invariant} if $s \in L$ implies $t \in L$,
for every forest~$t$ bisimilar to~$s$.
\begin{Thm}\label{Thm: bisimulation invariance}
A forest language $L \subseteq \bbF_0\Sigma$ is bisimulation-invariant if, and only if,
the syntactic algebra~$\frakS(L)$ satisfies the following equations\?:
\begin{alignat*}{-1}
  c + c &= c\,,\qquad
  & a(x_0 + x_0) &= a(x_0)\,,\\
  c + d &= d + c\,,\qquad
  & a(x_0 + x_1 + x_2 + x_3) &= a(x_0 + x_2 + x_1 + x_3)\,,
\end{alignat*}
for all $a \in S_1(L)$ and $c,d \in S_0(L)$.
\end{Thm}
\begin{proof}
Let $\eta : \bbF\Sigma \to \frakS(L)$ be the syntactic morphism mapping a forest to its
$\sim_L$-class.

$(\Rightarrow)$
Given elements $c,d \in S_0(L)$, we fix forests $s \in \eta^{-1}(c)$ and $t \in \eta^{-1}(d)$.
If $L$~is bisimulation-invariant, we have
\begin{align*}
  p[s] \in L \quad\iff\quad p[s + s] \in L
  \quad\qtextq{and}\quad
  p[s+t] \in L \quad\iff\quad p[t+s] \in L\,,
\end{align*}
for every context~$p$.
Consequently, $s \sim_L s + s$ and $s + t \sim_L t + s$, which implies that $c = c + c$
and $c + d = d + c$.

The remaining two equations are proved similarly.
Fix $a \in S_1(L)$ and $s \in \eta^{-1}(a)$.
Setting $s' := s(x_0 + x_0)$, bisimulation-invariance of~$L$ implies that
\begin{align*}
  p[s] \in L \quad\iff\quad p[s'] \in L\,,
  \quad\text{for every context } p\,.
\end{align*}
Consequently $s \sim_L s'$ and $a(x_0) = \eta(s) = \eta(s') = a(x_0 + x_0)$.

Similarly, for $t := s(x_0 + x_1 + x_2 + x_3)$ and $t' := s(x_0 + x_2 + x_1 + x_3)$, we have
\begin{align*}
  p[t] \in L \quad\iff\quad p[t'] \in L\,,
  \quad\text{for every context } p\,.
\end{align*}
Hence, $t \sim_L t'$ and $a(x_0 + x_1 + x_2 + x_3) = a(x_0 + x_2 + x_1 + x_3)$.

$(\Leftarrow)$
Suppose that $\frakS(L)$ satisfies the four equations above and let $s$~and~$s'$
be bisimilar forests.
We claim that $\eta(s) = \eta(s')$, which implies that $s \in L \Leftrightarrow s' \in L$.

Fix a bisimulation relation $Z \subseteq \dom(s) \times \dom(s')$. W.l.o.g.\ we may assume
that $Z$~only relates vertices on the same level of the respective forests and that it only
relates vertices whose predecessors are also related.
(If not, we can always remove the pairs not satisfying this condition without destroying the fact
that $Z$~is a bisimulation.)
Let $\approx$~be the equivalence relation on $\dom(s) \cup \dom(s')$ generated by~$Z$.

We will transform the forests $s$~and~$s'$ in several steps while preserving
their value under~$\eta$ until both forests are equal.
(Note that each of these steps necessarily modifies the given forest at every vertex.)
An example of this process can be found in Figure~\ref{Fig: bisim transformation}.
\begin{figure}
\centering
\includegraphics{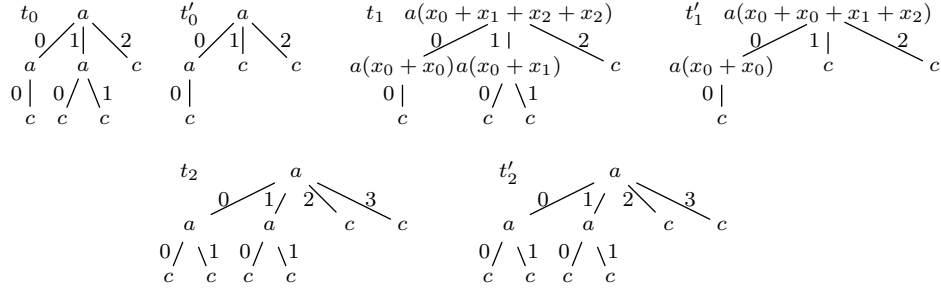}
\caption{Transforming bisimilar forests\label{Fig: bisim transformation}}
\end{figure}
The first step consists in translating the problem into the algebra~$\frakS(L)$.
We define two new forests $t_0,t'_0 \in \bbF_0S(L)$ with the same domains as, respectively,
$s$~and~$s'$ and the following labelling.
If $v \in \dom(s)$ has the $0$-successors $u_0,\dots,u_{n-1}$, we set
\begin{align*}
  t_0(v) := \eta(s(v))(x_0 +\dots+ x_{n-1})
\end{align*}
and we make $u_i$~an $i$-successor of~$v$ in~$t_0$.
We obtain~$t'_0$ from~$s'$ in the same way.
By associativity it follows that $\pi(t_0) = \eta(s)$ and $\pi(t'_0) = \eta(s')$.

Next we make the shapes of the forests $t_0$~and~$t'_0$ the same.
Let $t_1$~and~$t'_1$ be the forests with the same domains as $t_0$~and~$t'_0$ and
the following labelling.
For every vertex~$v$ of~$t_0$ with successors $u_0,\dots,u_{n-1}$ and labelling
\begin{align*}
  t_0(v) = a(x_0 +\dots+ x_{n-1})\,,
\end{align*}
we set
\begin{align*}
  t_1(v) := a(x_0 +\dots + x_0 +\dots+ x_{n-1}+\dots+x_{n-1})\,,
\end{align*}
where each variable~$x_i$ is repeated~$m_i$ times and the numbers~$m_i$
are determined as follows.
Let $M$~be some number such that, for every $i < n$,
no vertex $v' \approx v$ has at more than~$M$ successors~$u'$ with $u' \approx u_i$.
(Note that there are only finitely many such vertices.)
We choose the constants~$m_i$ such that
\begin{align*}
  \sum_{k \in U_i} m_k = M\,,
  \qtextq{where}
  U_i := \set{ k < n }{ u_k \approx u_i }\,.
\end{align*}
We obtain the forest~$t'_1$ in the same way from~$t'_0$.
By the top right equation in the statement of the theorem,
the value of the product is not affected by this modification.
Hence, $\pi(t_1) = \pi(t_0)$ and $\pi(t'_1) = \pi(t'_0)$.

Finally, let $t_2$~and~$t'_2$ be the unravelling of, respectively, $t_1$~and~$t'_1$, i.e.,
the forest where for every vertex~$v$ with successors $u_0,\dots,u_{n-1}$ and label
\begin{align*}
  t_1(v) = a(x_0 +\dots+ x_0 +\dots+ x_{n-1} +\dots+ x_{n-1})\,,
\end{align*}
we set
\begin{align*}
  t_2(v) := a(x_0 +\dots+ x_k +\dots+ x_l +\dots+ x_m)
\end{align*}
(where we number the variables from left-to-right, e.g.,
$a(x_0+x_0+x_1+x_2+x_2)$ becomes $a(x_0+x_1+x_2+x_3+x_4)$),
and we duplicate each attached subforest a corresponding number of times such that
the value of the product does not change.
We do the same for~$t'_2$.

We have arrived at a situation where, for each component~$r$ of the forests~$t_2$,
there is some component~$r'$ of~$t'_2$ that differs only in the ordering of successors,
but not in their number.
Consequently, there exists a bijection $\sigma : \dom(t) \to \dom(r')$ such that,
for a vertex~$v$ of~$r$ with successors $u_0,\dots,u_{n-1}$,
\begin{align*}
  r'(v) = r(v)(x_{\sigma_v(0)}+\dots+x_{\sigma_v(n-1)})\,,
\end{align*}
where the function $\sigma_v : [n] \to [n]$ is chosen such that
$\sigma(u_i)$ is the $\sigma_v(i)$-successor of~$\sigma(v)$.

Let $\hat r$~be the tree obtained from~$r$ as follows.
For a vertex~$v$ with successors $u_0,\dots,u_{n-1}$ and labelling
\begin{align*}
  r(v) = a(x_0+\dots+x_{n-1})\,,
\end{align*}
we set
\begin{align*}
  \hat r(v) := a(x_{\sigma_v(0)}+\dots+x_{\sigma_v(n-1)})\,,
\end{align*}
and we reorder the attached subtrees accordingly.
By associativity and the bottom right equation, this does not change the value of the product.
It follows that $\hat r = r'$.
Consequently, $\pi(r) = \pi(r')$.

We have shown that, for every component of~$t_0$ there is some component of~$t'_0$
with the same product. Therefore, we can write
\begin{align*}
  \pi(t_0) = a_0 +\dots+ a_{m-1}
  \qtextq{and}
  \pi(t'_0) = b_0 +\dots+ b_{n-1}
\end{align*}
where the sets $\{a_0,\dots,a_{m-1}\}$ and $\{b_0,\dots,b_{m-1}\}$ coincide.
Using the equations $c+c=c$ and $c+d=d+c$ we can therefore transform $\pi(t_0)$ into~$\pi(t'_0)$.
Consequently,
\begin{align*}
  \eta(s) = \pi(t_0) = \pi(t'_0) = \eta(s')\,.
\end{align*}
As $\eta$~recognises~$L$ it follows that $s \in L \Leftrightarrow s' \in L$,
as desired.
\end{proof}

Note that we immediately obtain a decision procedure for bisimulation-invariance
from this theorem,
since we can compute the syntactic algebra and check whether it satisfies the given
set of equations.
\begin{Cor}
It is decidable whether a given regular language~$L$ is bisimulation-invariant.
\end{Cor}

\section{The Logic cEF}   
\label{Sect: cEF}

Let us now proceed to the main result of this article\?:
a~characterisation of the temporal logic $\cEF$.
For simplicity, the following definition of its semantics only considers
forests instead of arbitrary transition systems.
\begin{Def}
(a)
\emph{Counting $\EF$,} $\cEF$ for short,
has two kinds of formulae\?: \emph{tree formulae} and \emph{forest formulae,}
which are inductively defined as follows.
\begin{itemize}
\item Every forest formula is a finite boolean combination of formulae of the form
  $\sfE_k\varphi$ where $k$~is a positive integer and $\varphi$~a tree formula.
\item Every tree formula is a finite boolean combination of (i)~forest formulae and
  (ii)~formulae of the form~$P_a$, for $a \in \Sigma$.
\end{itemize}

To define the semantics we introduce a satisfaction relation~$\models_\rmf$ for forest formulae and
one~$\models_\rmt$ for tree formulae.
In both cases boolean combinations are defined in the usual way.
For a tree~$t$, we define
\begin{align*}
  t \models_\rmt P_a     &\quad\defiff\quad \text{the root of } t \text{ has label } a\,, \\
  t \models_\rmt \varphi &\quad\defiff\quad
    t' \models_\rmf \varphi\,,
    \begin{aligned}[t]
      &\quad\text{for a forest formula } \varphi\,, \text{ where $t'$~denotes the successor} \\
      &\quad\text{forest of the root of~$t$\,.}
    \end{aligned}
\end{align*}
For a forest~$s$, we define
\begin{align*}
  s \models_\rmf \sfE_k\varphi
  \quad\defiff\quad
  &\text{there exist at least~$k$ vertices~$v$, distinct from the roots, such that} \\
  &s|_v \models \varphi\,.
\end{align*}

(b) For $k,m < \omega$, we denote by $\cEF_k$ the fragment of $\cEF$ that
uses only operators~$\sfE_l$ where $l \leq k$, and
$\cEF_k^m$ is the fragment of $\cEF_k$ where
the nesting depth of the operators~$\sfE_l$ is restricted to~$m$.
For $k = 1$, we set $\EF := \cEF_1$ and $\EF^m := \cEF_1^m$.
\markenddef
\end{Def}

The following is our main theorem.
Before giving the statement a few technical remarks are in order.
In the equations below we make use of the \emph{$\omega$-power} $a^\omega$ of an element
$a \in A_1$ (which is the infinite vertical product $aaa\dots$),
and the \emph{idempotent power}~$a^\pi$ (which is the defined as $a^\pi = a^n$ for
the minimal number~$n$ with $a^na^n = a^n$).
For the horizontal semigroup we use multiplicative notation instead\?: $n \times a$ for
$a +\dots+ a$ and $\pi \times a$ for $n \times a$ with $n$~as above.

When writing an $\omega$-power of an element of arity greater than one, we need to specify with
respect to which variable we take the power. We use the notation~$a^{\omega_i}$ to indicate
that the variable~$x_i$ should be used.
Note that, when using several $\omega$-powers like in $(a(x_0, (b(x_0,x_1))^{\omega_1}))^{\omega_0}$,
the intermediate term after resolving the inner power can be a forest with infinitely many
occurrences of the variable~$x_0$.
But after resolving the outer $\omega$-power, we obtain a forest without variables, i.e.,
a proper element of~$\bbF_0 A$. Consequently, the equations below are all well-defined.
Finally, to keep notation light we will frequently write~$x$ instead of~$x_0$,
if this is the only variable present.
\begin{Thm}\label{Thm: cEF-definability}
A forest language $L \subseteq \bbF_0\Sigma$ is definable in the logic~$\cEF_k$ if, and only if,
the syntactic algebra~$\frakS(L)$ satisfies the following equations\?:
\begin{alignat*}{-1}
  c + d &= d + c &\qquad\qquad
  (a(x) + b(x))^\omega &= (ab(x))^\omega \\
  (ab)^\pi &= b(ab)^\pi &\qquad\qquad
  (a(x) + c)^\omega &= (a(x+c))^\omega \\
  a^\omega + a^\omega &= a^\omega &\qquad\qquad
  (a(x + c + c))^\omega &= (a(x + c))^\omega \\
  (abb')^\omega &= (ab'b)^\omega &\qquad\qquad
  \bigl[a(b(x_0,x_1))^{\omega_1}\bigr]^{\omega_0} &= [ab(x_0,x_0)]^{\omega_0} \\
  (aab)^\omega &= (ab)^\omega &\qquad\qquad
  [a(x + bc + c)]^\omega &= [a(x + bc)]^\omega \\[1ex]
  &\mkern-36mu\rlap{\begin{aligned}
           a_n(c,\dots,c) + (k-n)\times c &= a_n(c,\dots,c) + (k-n+1)\times c\,, \\
          [a(x + (a(k\times x))^\pi(c))]^\omega &= k\times (a(k\times x))^\pi(c)
   \end{aligned}}
\end{alignat*}
for all $a,b,b' \in S_1(L)$, $c,d \in S_0(L)$, $a_n \in S_n(L)$, and $n \leq k$.
\end{Thm}
No attempt was made to simplify the above axioms.
While having a simpler description would of course be nice, the importance of this result
lies in the facts that (i)~an equational axiomatisation exists\?;
that (ii)~the equations can be checked algorithmically\?;
and (iii)~that our framework was sufficient to derive them.

We defer the proof to Section~\ref{Sect: cEF proof}.
Let us concentrate on some of the consequences first.

\begin{Cor}
For fixed~$k$,
it is decidable whether a given regular language~$L$ is $\cEF_k$-definable.
\end{Cor}

For the logic $\cEF$, where the value of~$k$ is not bounded, a similar result can
now be derived as a simple corollary.
The basic argument is contained in the following lemma.
\begin{Lem}
Given a forest algebra\/~$\frakA$ that is generated by $A_0 \cup A_1$,
we can compute a number~$K$ such that,
if\/ $\frakA$~satisfies the equations of Theorem~\ref{Thm: cEF-definability}
for some value of~$k$, it satisfies them for $k = K$.
\end{Lem}
\begin{proof}
Set $K := m_0^{2m_1} + m_0$ where $m_0 := \abs{A_0}$ and $m_1 := \abs{A_1}$.
By assumption there is some number~$k$ for which $\frakA$~satisfies the equations of
Theorem~\ref{Thm: cEF-definability}.
W.l.o.g.\ we may assume that $k \geq K$.
The only two equations depending on~$k$ are
\begin{enumerate}[label=\textup{(\arabic*)${}_k$}, leftmargin=3em]
\item $a_n(c,\dots,c) + (k-n)\times c = a_n(c,\dots,c) + (k-n+1)\times c$
\item $[a(x + (a(k\times x))^\pi(c))]^\omega = k\times (a(k\times x))^\pi(c)$
\end{enumerate}
We have to show that $\frakA$~also satisfies $(1)_K$ and~$(2)_K$.

For $(2)_K$,
note that $k \geq K \geq \abs{A_0}$ implies that
$K \times c = \pi\times c = k\times c$, for all $c \in A_0$.
Consequently,
\begin{align*}
  a(K\times x)(c) = a(k\times x)(c)
  \qtextq{and, therefore,}
  (a(K\times x))^\pi(c) = (a(k\times x))^\pi(c)\,.
\end{align*}
This implies the claim.

For $(1)_K$, fix $a \in A_n$ and $c \in A_0$.
If $n \leq K - m_0$, then $K-n \geq m_0 = \abs{A_0}$ implies that $(K-n)\times c = \pi\times c$.
Consequently,
\begin{align*}
  a(c,\dots,c) + (K-n)\times c
  = a(c,\dots,c) + \pi \times c
  = a(c,\dots,c) + \pi \times c + c
\end{align*}
and we are done.
Thus, we may assume that $n > K - m_0 = m_0^{2m_1}$.
As $\frakA$~is generated by $A_0 \cup A_1$, there exists some forest $s \in \bbF_i(A_0 \cup A_1)$
with $\pi(s) = a$.
We distinguish several cases.

If some of the variables $x_0,\dots,x_{n-1}$ does not appear in~$s$,
we can use~$(1)_k$ to show that
\begin{align*}
  a(c,\dots,c,\dots,c) + (K-n)\times c
  &= a(c,\dots,c+\dots+c,\dots c) + (K-n)\times c \\
  &= a(c,\dots,k\times c,\dots, c) + (K-n)\times c \\
  &= a(c,\dots,k\times c,\dots, c) + (K-n)\times c + c\,.
\end{align*}

Next, suppose that $s$~is highly branching in the sense that it has the form
\begin{align*}
  s = r(t_0 +\dots+ t_{m_0^2-1})
\end{align*}
where each subterm~$t_i$ contains some variable.
Then there are indices $i_0 < \dots < i_{m_0-1}$ such that
$\pi(t_{i_0}(\bar c)) = \dots = \pi(t_{i_{m_0-1}}(\bar c))$
(where $\bar c$~denotes as many copies of~$c$ as appear in the respective term).
Hence, $(1)_k$~again implies that
\begin{align*}
  a(\bar c) + (K-n)\times c
  &= \pi(s(\bar c)) + (K-n)\times c \\
  &= \pi\bigl(r\bigl(t_0(\bar c) +\dots+ t_{m_0^2-1}(\bar c)\bigr)\bigr) + (K-n)\times c \\
  &= \pi\bigl(r\bigl(t_0(\bar c) +\dots+ t_{m_0^2-1}(\bar c) + k \times t_{i_0}(\bar c)\bigr)\bigr)
       + (K-n)\times c \\
  &= a(\bar c) + (K-n)\times c + c\,.
\end{align*}

Note that a tree of height $h := m_1$ where every vertex has at most $d := m_0^2$ successors
has at most $d^h = m_0^{2m_1}$ leaves.
Hence, if $s$~is not highly branching in the sense above, the fact that it contains
$n > m_0^{2m_1}$ variables implies that there must be a chain $v_0 \prec\dots\prec v_{m_1}$
of vertices such that, for every $i < m_1$, there is some leaf~$u$ labelled by a variable with
$v_{i-1} \prec u$ and $v_i \npreceq u$.
(For $i = 0$, we omit the first condition.)
Hence, we can decompose~$s$ as
\begin{align*}
  s(\bar c) = r_0(\bar c,r_1(\bar c, \dots r_{m_1}(\bar c)))\,,
\end{align*}
and there are two indices $i < j$ such that
\begin{align*}
  \pi(r_0(\bar c,\dots r_i(\bar c,x))) = \pi(r_0(\bar c,\dots r_j(\bar c,x)))\,.
\end{align*}
Consequently, we can use pumping to obtain a term
\begin{align*}
  \pi(s(\bar c))
  = \pi\bigl(r_0(\bar c,\dots, r_i(\bar c,x))\bigl[r_{i+1}(\bar c,\dots, r_j(\bar c,x))\bigr]^k
            r_{j+1}(\bar c,\dots, r_{m_1}(\bar c))\bigr)
\end{align*}
which contains at least~$k$ occurrences of~$c$,
and the claim follows again by~$(1)_k$.
\end{proof}

According to this lemma, we can check for $\cEF$-definability of a language~$L$,
by computing its syntactic algebra~$\frakS(L)$, the associated constant~$K$,
and then checking the equations for $k = K$.
\begin{Cor}
It is decidable whether a given regular language~$L$ is $\cEF$-definable.
\end{Cor}

When taking the special case of $k = 1$ in Theorem~\ref{Thm: cEF-definability},
we obtain the following characterisation of $\EF$-definability.
\begin{Thm}\label{Thm: EF-definability}
A forest language $L \subseteq \bbF_0\Sigma$ is definable in the logic~$\EF$ if, and only if,
the syntactic algebra~$\frakS(L)$ satisfies the following equations\?:
\begin{alignat*}{-1}
  c + d &= d + c &\qquad\qquad
  (a(x) + b(x))^\omega &= (ab(x))^\omega \\
  (ab)^\pi &= b(ab)^\pi &\qquad\qquad
  (a(x) + c)^\omega &= (a(x+c))^\omega \\
  (abb')^\omega &= (ab'b)^\omega &\qquad\qquad
  (a(x + c + c))^\omega &= (a(x + c))^\omega \\
  (aab)^\omega &= (ab)^\omega &\qquad\qquad
  \bigl[a(b(x_0,x_1))^{\omega_1}\bigr]^{\omega_0} &= [ab(x_0,x_0)]^{\omega_0} \\[1ex]
  &\mkern-36mu\rlap{\begin{aligned}
          &ac = ac+c \qquad c = c+c \qquad [a(x + a^\pi c)]^\omega = a^\pi c\,,
   \end{aligned}}
\end{alignat*}
for all $a,b,b' \in S_1(L)$ and $c,d \in S_0(L)$.
\end{Thm}
\begin{Cor}
It is decidable whether a given regular language~$L$ is $\EF$-definable.
\end{Cor}

\section{The proof of Theorem~\ref{Thm: cEF-definability}}   
\label{Sect: cEF proof}

For the proof of Theorem~\ref{Thm: cEF-definability}, we need to set up a bit of machinery.
We start by defining the suitable notion of bisimulation for~$\cEF_k$.
The difference to the standard notion is that we use reachability instead of the edge relation
and that we also have to preserve the number of reachable positions.
\begin{Def}
Let $m,k < \omega$.

(a) For trees $s,t \in \bbF\Sigma$, we define
\begin{alignat*}{-1}
  s &\approx_k^0 t   &&\quad\defiff\quad &&\text{the roots of $s$ and $t$ have the same label} \\
  s &\approx_k^{m+1} t &&\quad\defiff\quad
     &&\text{the roots of $s$ and $t$ have the same label}\,, \\
  &&&&&\text{for every $k$-tuple } \bar x \text{ in } \dom(s) \text{ not containing the root,
             there is} \\
  &&&&&\quad\text{some $k$-tuple } \bar y \text{ in } \dom(t) \text{ not containing the root
             such that}\\
  &&&&&\qquad s|_{x_i} \approx_k^m t|_{y_i} \qtextq{and} x_i = x_j \Leftrightarrow y_i = y_j\,,
       \quad\text{for all } i,j < k \text{ and,} \\
  &&&&&\text{for every $k$-tuple } \bar y \text{ in } \dom(t) \text{ not containing the root,
             there is} \\
  &&&&&\quad\text{some $k$-tuple } \bar x \text{ in } \dom(s) \text{ not containing the root
             such that}\\
  &&&&&\qquad s|_{x_i} \approx_k^m t|_{y_i} \qtextq{and} x_i = x_j \Leftrightarrow y_i = y_j\,,
       \quad\text{for all } i,j < k\,.
\end{alignat*}
To simplify notation, we will frequently write $x \approx_k^m y$ for vertices $x$~and~$y$
instead of the more cumbersome $s|_x \approx_k^m t|_y$.

(b) For forests $s,t \in \bbF\Sigma$ with possibly several components, we set
\begin{alignat*}{-1}
  s &\sim_k^{m+1} t &\quad\defiff\quad&
     &&\text{for every $k$-tuple } \bar x \text{ in } s \text{ there is some $k$-tuple }
          \bar y \text{ in } t \text{ such that} \\
  &&&&&\qquad s|_{x_i} \approx_k^m t|_{y_i} \qtextq{and} x_i = x_j \Leftrightarrow y_i = y_j\,,
       \quad\text{for all } i,j < k \text{ and,} \\
  &&&&&\text{for every $k$-tuple } \bar y \text{ in } t \text{ there is some $k$-tuple }
         \bar x \text{ in } s \text{ such that}\\
  &&&&&\qquad s|_{x_i} \approx_k^m t|_{y_i} \qtextq{and} x_i = x_j \Leftrightarrow y_i = y_j\,,
       \quad\text{for all } i,j < k\,.
\end{alignat*}
\upqed
\markenddef
\end{Def}

Let us show that this notion of bisimulation captures the expressive power of~$\cEF$.
The proof is mostly standard. We start by introducing the following notion of a type.
\begin{Def}
(a) We define the \emph{type} $\tp_k^m(s)$ of a tree $s \in \bbF\Sigma$ by
\begin{align*}
  \tp_k^0(s) &:= a \\
  \tp_k^{m+1}(s) &:= \langle a, \theta_s\rangle
\end{align*}
where $a$~is the label at the root of~$s$ and
\begin{align*}
  \theta_s := \biglset \langle l,\sigma\rangle \bigmset {}
                     & l \leq k\,,\ x_0,\dots,x_{l-1} \in \dom(s) \text{ distinct, not equal to the root}\,,\ \\
                     & \sigma = \tp_k^m(s|_{x_0}) = \dots = \tp_k^m(s|_{x_{l-1}}) \bigrset\,.
\end{align*}

(b) For an arbitrary forest $s \in \bbF\Sigma$, we set
\begin{align*}
  \Tp_k^{m+1}(s) &:= \theta_s\,,
\end{align*}
where
\begin{align*}
  \theta_s := \biglset \langle l,\sigma\rangle \bigmset {}
                     & l \leq k\,,\ x_0,\dots,x_{l-1} \in \dom(s) \text{ distinct}\,,\ \\
                     & \sigma = \tp_k^m(s|_{x_0}) = \dots = \tp_k^m(s|_{x_{l-1}}) \bigrset\,.
\end{align*}
\upqed
\markenddef
\end{Def}

A standard proof establishes the following equivalences.
\begin{Lem}\label{Lem: basic property of simtr}
Let $k,m < \omega$.
\begin{enuma}
\item For trees $s,t \in \bbF_0\Sigma$, the following statements are equivalent.
  \begin{enum1}
  \item $s \approx_k^m t$
  \item $\tp_k^m(s) = \tp_k^m(t)$
  \item $s \models \varphi \Leftrightarrow t \models \varphi\,, \text{ for all } \varphi \in \cEF_k^m\,.$
  \end{enum1}
\item For arbitrary forests $s,t \in \bbF_0\Sigma$, the following statements are equivalent.
  \begin{enum1}
  \item $s \sim_k^m t$
  \item $\Tp_k^m(s) = \Tp_k^m(t)$
  \item $s \models \varphi \Leftrightarrow t \models \varphi\,, \text{ for all } \varphi \in \cEF_k^m\,.$
  \end{enum1}
\end{enuma}
\end{Lem}
\begin{proof}
(a)
(2)~$\Rightarrow$~(1) follows by a straightforward induction on~$m$ and
(1)~$\Rightarrow$~(3) by induction on~$\varphi$.
For (3)~$\Rightarrow$~(2) it is sufficient to show that,
for every type~$\tau$, there exists a formula $\chi_\tau \in \EF_k^m$ such that
\begin{align*}
  s \models \chi_\tau \quad\iff\quad \tp_k^m(s) = \tau\,, \quad\text{for every tree } s\,.
\end{align*}
We proceed by induction on~$m$.
If $m = 0$, the type~$\tau$ is of the form $a \in \Sigma$. Hence, we can set
$\chi_\tau := P_a$.
If $m > 0$, then $\tau = \langle a,\theta\rangle$ for some $a \in \Sigma$ and some
set $\theta$~of types of lower rank. We can set
\begin{align*}
  \chi_\tau := P_a \land \Land_{\langle l,\sigma\rangle \in \theta} \sfE_l\chi_\sigma
                   \land \Land_{\langle l,\sigma\rangle \notin \theta} \neg\sfE_l\chi_\sigma\,.
\end{align*}

(b) is proved in the same way.
\end{proof}

\begin{Cor}\label{Cor: cEF-definable}
A language $L \subseteq \bbF\Sigma$ is $\cEF_k^m$-definable if, and only if,
it is regular and satisfies
\begin{align*}
  s \sim_k^m t \qtextq{implies} s \in L \Leftrightarrow t \in L\,,
  \quad\text{for all regular forests } s,t \in \bbF_0\Sigma\,.
\end{align*}
\end{Cor}
\begin{proof}
$(\Rightarrow)$ follows by the implication $(1) \Rightarrow (3)$
of Lemma~\ref{Lem: basic property of simtr}.

$(\Leftarrow)$ Set
\begin{align*}
  \varphi := \Lor {\bigset{ \chi_\tau }{ \tau = \Tp_k^m(s) \text{ for some regular forest } s \in L }}\,,
\end{align*}
where $\chi_\tau$~are the formulae from the proof of Lemma~\ref{Lem: basic property of simtr}.
For a regular forest $t \in \bbF_0\Sigma$, it follows that
\begin{align*}
  t \models \varphi
  &\quad\iff\quad
  \Tp_k^m(t) = \Tp_k^m(s)\,, \quad\text{for some regular forest } s \in L\,, \\
  &\quad\iff\quad
  t \sim_k^m s\,, \quad\text{for some regular forest } s \in L\,, \\
  &\quad\iff\quad
  t \in L\,.
\end{align*}
Let $K$~be the language defined by~$\varphi$.
Since $L$~and~$K$ are both regular languages that contain the same regular forests,
it follows that $L = K$. Thus, $L$~is $\cEF_k^m$-definable.
\end{proof}

We want to show that an algebra recognises $\cEF_k$-definable languages if, and only if,
it satisfies the following equations.
\begin{Def}\label{Def: algebra for cEF}
(a) A forest algebra~$\frakA$ is an \emph{algebra for $\cEF_k$} if it is finitary,
generated by $A_0 \cup A_1$, and satisfies the following equations.
\begin{enumerate}[label=\textup{(G\arabic*)\hphantom{${}_k$}}, leftmargin=3em]
\item[(G1)${}_k$] $a_n(c,\dots,c) + (k-n)\times c = a_n(c,\dots,c) + (k-n+1)\times c$
\addtocounter{enumi}{1}%
\item $(ab)^\pi = b(ab)^\pi$
\item $a^\omega + a^\omega = a^\omega$
\item $c + d = d + c$
\item $(a(x) + b(x))^\omega = (ab(x))^\omega$
\item $(a(x) + c)^\omega = (a(x+c))^\omega$
\item $(a(x + c + c))^\omega = (a(x + c))^\omega$
\item $\bigl[a(b(x_0,x_1))^{\omega_1}\bigr]^{\omega_0} = [ab(x_0,x_0)]^{\omega_0}$
\item $(abb')^\omega = (ab'b)^\omega$
\item $(aab)^\omega = (ab)^\omega$
\item $[a(x + bc + c)]^\omega = [a(x + bc)]^\omega$
\item[(G12)${}_k$] $[a(x + (a(k\times x))^\pi(c))]^\omega = k\times (a(k\times x))^\pi(c)$
\end{enumerate}
where $a,b,b' \in A_1$, $c,d \in A_0$, $a_n \in A_n$, and $n \leq k$.

(b) A forest algebra~$\frakA$ is an \emph{algebra for $\cEF$} if it is an algebra for $\cEF_k$,
for some $k \geq 1$.
\markenddef
\end{Def}

In the proof that algebras for $\cEF$ recognise exactly the $\cEF$-definable languages,
we use one of the Green's relations (suitably modified for forest algebras).
\begin{Def}
Let $\frakA$~be a forest algebra. For $a,b \in A_0$, we define
\begin{align*}
  a \leq_\sfL b \quad\defiff\quad
  a = c(b) \qtextq{or} a = b + d\,,
  \quad\text{for some } c \in A_1\,,\ d \in A_0\,.
\end{align*}
\upqed
\markenddef
\end{Def}

\begin{Lem}\label{Lem: basic properties of leq_L}
Let\/ $\frakA$~be an algebra for $\cEF_k$.
\begin{enuma}
\item The relation $\leq_\sfL$~is antisymmetric.
\item For $a \in A_1\,,\ c \in A_0$, we have
  \begin{alignat*}{-1}
    c &= c + c  &&\qtextq{implies} &ac &= ac + c\,, \\
    c &= a(c,c) &&\qtextq{implies} &c &= c + c\,.
  \end{alignat*}
\end{enuma}
\end{Lem}
\begin{proof}
(a)
For a contradiction, suppose that there are elements $a \neq b$ with
$a \leq_\sfL b \leq_\sfL a$.
By definition, we can find elements $c$~and~$d$ such that
(1)~$a = c(b)$ or (2)~$a = b + c$, and (i)~$b = d(a)$ or (ii)~$b = a + d$.
We have thus to consider four cases.
In each of them we obtain a contradiction via (G1)${}_k$ or (G2).
\begin{align*}
  (1,i)\quad  a &= cb = cda = (cd)^\pi(a) = d(cd)^\pi(a) = da = b\,. \\
  (1,ii)\quad a &= cb = c(a+d) = (c(x+d))^\pi(a) = (c(x+d))^\pi(a) + d = a+d = b\,. \\
  (2,i)\quad  b &= da = d(b+c) = (d(x+c))^\pi(b) = (d(x+c))^\pi(b) + c = b+c = a\,. \\
  (2,ii)\quad a &= b + c = a + d + c = a + k\times(d+c) = a + k\times(d+c) + d = a + d = b\,.
\end{align*}

(b) By (G1)${}_k$ we have
\begin{alignat*}{-1}
  c &= c + c &&\qtextq{implies}
    &ac &= a(c+c) = a(k\times c) = a(k\times c)+c = ac+c\,, \\
  c &= a(c,c) &&\qtextq{implies}
    &c &= a(c,c) = (a(x,c))^\pi(c) = (a(x,c))^\pi(c) + c = c + c\,.
\end{alignat*}
\upqed
\end{proof}

\begin{figure}
\centering
\includegraphics{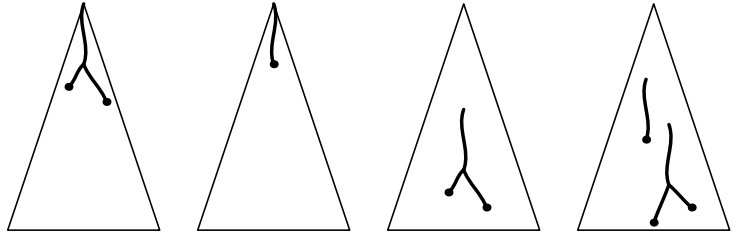}
%
%
%
%
%
%
%
%
%
%
\caption{A forest~$s$ with a convex set~$U$ (in bold) that has three close $U$-ends (on the left)
and five far ones (on the right). The height is $h(s,U) = 2$.\label{Fig:U-ends}}
\end{figure}

Let us take a look at the following situation (see Figure~\ref{Fig:U-ends}).
Let $s$~be a forest and $U$~a set of vertices.
We assume that $U$~is \emph{convex} in the sense that
$u \preceq v \preceq w$ and $u,w \in U$ implies $v \in U$
(where $\preceq$~denotes the forest order).
We call the maximal elements (w.r.t.~$\preceq$) of~$U$ the \emph{$U$-ends.}
An $U$-end~$u$ is \emph{close} if $u' \in U$, for all $u' \preceq u$.
Otherwise, it is \emph{far.}
We would like to know how many of the $U$-ends are close.

\begin{Lem}\label{Lem: number of close ends}
Let $m \geq 0$ and $k \geq 1$,
let $s \sim_k^{m+k+2} t$ be two forests, $U \subseteq \dom(s)$ a convex set that is
closed under~$\approx_k^m$, and set
\begin{align*}
  V := \set{ v \in \dom(t) }{ u \approx_k^m v \text{ for some } u \in U }\,.
\end{align*}
\begin{enuma}
\item $V$~is convex and closed under~$\approx_k^m$.
\item The numbers of ends of $U$~and~$V$ are the same, or both numbers are at least~$k$.
\item If $U$ has less than~$k$ ends, then $U$~is finite if, and only if, $V$~is finite.
\item If $U$~is finite and has less than~$k$ ends,
  then $U$~and~$V$ have the same numbers of close ends and of far ones.
\end{enuma}
\end{Lem}
\begin{proof}
(a)
If $V$~is not convex, there are vertices $v \prec v' \prec v''$ of~$t$ with
$v,v'' \in V$ and $v' \notin V$. Fix vertices $u \prec u' \prec u''$ with
$u \approx_k^{m+2} v$, $u' \approx_k^{m+1} v'$, and $u'' \approx_k^m v''$.
By definition of~$V$, we have $u,u'' \in U$ and $u' \notin U$.
This contradicts the fact that $U$~is convex.

To see that $V$~is closed under~$\approx_k^m$, suppose that $v \in V$ and $v \approx_k^m v'$.
By definition of~$V$, there is some $u \in U$ with $u \approx_k^m v$.
Hence, $u \approx_k^m v \approx_k^m v'$. As $\approx_k^m$~is transitive, this implies that $v' \in V$.

(b)
For a contradiction, suppose that $U$~has $n < k$ ends while $V$~has more than~$n$ ends.
(By~(a), the other case follows by symmetry.)
Choose $n+1$ ends $v_0,\dots,v_n \in V$. Since $s \approx_k^{m+2} t$, there are vertices
$u_0,\dots,u_n$ in~$s$ with $u_i \approx_k^{m+1} v_i$.
By definition of~$V$, we have $u_i \in U$.
By assumption, there is some index~$j$ such that $u_j$~is not an end.
Hence, we can find a vertex $u' \succ u_j$ with $u' \in U$.
Fix a vertex $v' \succ v_j$ of~$t$ with $u' \approx_k^m v'$.
Then $v' \in V$ and $v_j$~is not an end. A~contradiction.

(c)
For a contradiction, suppose that $U$~is finite, but $V$~is not.
(The other case follows again by symmetry.)
By~(b), $V$~has only finitely many ends.
Hence, there is some element $v \in V$ such that $v \npreceq v'$ for every end~$v'$ of~$V$.
Since $s \approx_k^{m+3} t$, we can find a vertex~$u$ of~$s$ with $u \approx_k^{m+2} v$.
This implies that $u \in U$. As $U$~is finite, we can find some end~$u'$ of~$U$
with $u \preceq u'$. Fix some $v' \succeq v$ with $u' \approx_k^{m+1} v'$.
Then $u' \in U$ implies $v' \in V$.
By choice of~$v$, there is some $v'' \succ v'$ with $v'' \in V$.
Choose $u'' \succ u'$ with $u'' \approx_k^m v''$.
By choice of~$u'$, we have $u'' \notin U$. This contradicts the fact that $v'' \in V$.

(d)
By~(b), we only need to prove that the number of close ends is the same.
Let $\hat U$~and~$\hat V$ be the sets of $U$-ends and $V$-ends, respectively.
We denote by $N(s,U)$ the number of close $U$-ends and
by $F(s,U)$ the set of all proper subforests~$s'$ of~$s$ that are attached to some
vertex~$v$ that does not belong to~$U$ but where at least one root belongs to~$U$.
(A~forest~$s'$ is a \emph{proper subforest} of~$s$ attached at~$v$ if $s'$~can be obtained from
the subtree~$s|_v$ by removing the root~$v$.)
We define the following equivalence relation.
\begin{alignat*}{-1}
  \langle s,U\rangle &\asymp_0 \langle t,V\rangle
  &&\quad\defiff\quad
  &&N(s,U) = N(t,V)\,, \\[1ex]
  \langle s,U\rangle &\asymp_{i+1} \langle t,V\rangle
  &&\quad\defiff\quad
  &&N(s,U) = N(t,V) \text{ and} \\
  &&&&&\#_\tau(s,U) = \#_\tau(t,V)\,, \text{ for every $\asymp_i$-class~$\tau$,}
\end{alignat*}
where $\#_\tau(s,U)$ denotes the number of subforests $s' \in F(s,U)$ that belong to the
class~$\tau$.

We define the \emph{$U$-height} of~$s$ by
\begin{align*}
  h(s,U) := \begin{cases}
              0 &\text{if } F(s;U) = \emptyset \\
              1 + \max {\set{ h(s',U) }{ s' \in F(s,U) }} &\text{otherwise.}
            \end{cases}
\end{align*}

By induction on~$l$, we will prove the following claim\?:
\begin{align*}
  \llap{(*)\ }
  s \sim_k^{m+l+2} t
  \qtextq{and}
  h(s,U) \leq l
  \qtextq{implies}
  h(s,U) = h(t,V)
  \qtextq{and}
  \langle s,U\rangle \asymp_l \langle t,V\rangle\,.
\end{align*}
As $h(s,U) \leq \abs{\hat U} < k$, it then follows that
$\langle s,U\rangle \asymp_k \langle t,V\rangle$.
In particular, $N(s,U) = N(t,V)$, as desired.

It thus remains to prove~$(*)$.
First, consider the case where $l = 0$.
If $h(t,V) > 0$, there is some $V$-end~$v$ that is not close.
Fix some vertex $v' \prec v$ with $v' \notin V$.
Since $s \sim_k^{m+2} t$, we can find vertices $u' \prec u$ of~$s$ with
$u' \approx_k^{m+1} v'$ and $u \approx_k^m v$.
By definition of~$V$, it follows that $u' \notin U$ and $u \in U$.
As $U$~is finite, we can find some $U$-end $w \succeq u$. But $u' \prec u \preceq w$
implies that $w$~is not close. Hence, $h(s,U) > 0$. A~contradiction.

For the second part, suppose that $\langle s,U\rangle \nasymp_0 \langle t,V\rangle$,
that is, $N(s,U) \neq N(t,V)$.
By symmetry, we may assume that $m := N(s,U) < N(t,v)$.
Pick $m+1$ distinct close $V$-ends $v_0,\dots,v_m$.
Since $m+1 \leq k$ and $s \sim_k^{m+2} t$, there are elements $u_0,\dots,u_m \in \dom(s)$
with $u_i \approx_k^{m+1} v_i$.
There must be some index~$j$ such that $u_j$~is not a close $U$-end.
As $U$~is closed under~$\approx_k^m$ and $u_j \approx_k^m v_j \approx_k^m u$, for some $u \in U$,
it follows that $u_j \in U$.
Furthermore, $u_j \approx_k^{m+1} v_j$ and the fact that $v_j$~is a $V$-end
implies that $u' \notin U$, for all $u' \succ u_j$.
Thus, $u_j$~is a $U$-end.
But $h(s,U) = 0$ implies that all $U$-ends of~$s$ are close. A~contradiction.

For the inductive step, suppose that
$s \sim_k^{m+(l+1)+2} t$ holds but we have $h(s,U) \neq h(t,V)$ or
$\langle s,U\rangle \nasymp_{l+1} \langle t,V\rangle$.
We distinguish several cases.

(i) Suppose that $h(s,U) > h(t,V)$.
By definition of~$h$, there is a subforest $s' \in F(s,U)$ with $h(s',U) = h(s,U) - 1$.
Then there is some subforest~$t'$ of~$t$ with $s' \sim_k^{m+l+2} t'$.
By inductive hypothesis it follows that
\begin{align*}
  h(s,U) = h(s',U) + 1 = h(t',V) + 1 < h(t,V) + 1 \leq h(s,U)\,.
\end{align*}
A~contradiction.

(ii) Suppose that $h(s,U) < h(t,V)$.
By definition of~$h$, there is a subforest $t' \in F(t,V)$ with $h(t',V) = h(t,V) - 1$.
Fix a subforest~$s'$ of~$s$ with $s' \sim_k^{m+l+2} t'$.
By inductive hypothesis, it follows that
\begin{align*}
  h(s,U) > h(s',U) = h(t',V) = h(t,V) - 1 \geq h(s,U)\,.
\end{align*}
A~contradiction.

(iii) Suppose that $N(s,U) \neq N(t,v)$ and there is no $\asymp_l$-class~$\tau$
with $\#_\tau(s,U) \neq \#_\tau(t,V)$.
Then we have $\abs{\hat U} - N(s,U) = \abs{\hat V} - N(t,V)$.
Since $\abs{\hat U} = \abs{\hat V}$ it follows that $N(s,U) = N(t,V)$.
A~contradiction.

(iv) Finally, suppose that there is some $\asymp_l$-class~$\tau$
with $\#_\tau(s,U) \neq \#_\tau(t,V)$.
By symmetry, we may assume that $m := \#_\tau(s,U) < \#_\tau(t,V)$.
We choose $m+1$ vertices $v_0,\dots,v_m$ of~$t$ such that the attached subforests have class~$\tau$.
Since $s \sim_k^{m+(l+1)+2} t$ and $m+1 \leq k$, there are vertices $u_0,\dots,u_m$ of~$s$ such that
$u_i \sim_k^{m+l+2} v_i$, for all $i \leq m$.
Let $s_i$~be the subforest of~$s$ attached to~$u_i$, and $t_i$~the subforest of~$t$ attached
to~$v_i$.
By inductive hypothesis, it follows that $s_i \asymp_l t_i$, for $i \leq m$.
Thus, $s$~has at least $m+1$ different subforest in the class~$\tau$.
A~contradiction.
\end{proof}
\begin{Cor}\label{Cor: subtrees with value c}
Let $s \sim_k^{m+k+2} t$ be forests such that, for every $c \in A_0$, the sets
\begin{align*}
  U_c := \set{ x \in \dom(s) }{ \pi(s|_x) = c }
  \qtextq{and}
  V_c := \set{ y \in \dom(t) }{ \pi(t|_y) = c }
\end{align*}
are convex and closed under~$\approx_k^m$.
Then $\pi(s) = \pi(t)$.
\end{Cor}
\begin{proof}
Suppose that $s = s_0 +\dots+ s_{l-1}$ and $t = t_0 +\dots+ t_{n-1}$, for trees $s_i$~and~$t_i$.
It is sufficient to show that, for every $c \in A_0$ such that the number of components~$s_i$
with $\pi(s_i) = c$ is different from the number of~$t_i$ with $\pi(t_i) = c$, we have
\begin{align*}
  \pi(s) = \pi(s) + \pi\times c
  \qtextq{and}
  \pi(t) = \pi(t) + \pi\times c\,.
\end{align*}
Adding enough terms~$c$ to $\pi(s) = \pi(s_0) +\dots+ \pi(s_{l-1})$ and
$\pi(t) = \pi(t_0) +\dots+ \pi(t_{n-1})$ it then follows that $\pi(s) = \pi(t)$.

Hence, fix such an element~$c$.
By Lemma~\ref{Lem: number of close ends}, we obtain one of the following cases.

(i) $U$~and~$V$ both have at least~$k$ ends. Then they contain an antichain of size~$k$.
and we can write~$s$ as $r(s'_0,\dots,s'_{k-1})$ with $\pi(s'_i) = c$.
Hence, it follows by (G1)${}_k$ that
\begin{align*}
  \pi(s) = \pi(r)(c,\dots,c) = \pi(r)(c,\dots,c) + \pi\times c
  = \pi(s) + \pi\times c\,.
\end{align*}
For $t$~it follows in the same way that
\begin{align*}
  \pi(t) = \pi(t) + \pi\times c\,.
\end{align*}

(ii) Both $U$~and~$V$ are infinite, but each has less than~$k$ ends.
Then they contain an infinite chain and
we can use Ramsey's Theorem (or the fact that $s$~is regular)
to write $\pi(s)$ as $a'e^\omega$ where $ec = c = e^\omega$.
By (G3) and (G1)${}_k$ it follows that
\begin{align*}
  \pi(s) &= a'e^\omega = a'(e^\omega +\dots+ e^\omega) = a'(c +\dots+ c) \\
         &= a'(c +\dots+ c) + \pi\times c \\
         &= \pi(s) + \pi\times c\,.
\end{align*}
For~$t$, we similarly obtain
\begin{align*}
  \pi(t) = \pi(t) + \pi\times c\,.
\end{align*}

(iii) The last remaining case is where both $U$~and~$V$ are finite and they have the same number
of close ends. Then the number of indices~$i$ with $\pi(s_i) = c$ would be the same as the number
of~$i$ with $\pi(t_i) = c$, in contradiction to our choice of~$c$.
\end{proof}

Bevor presenting our main technical result,
let us quickly recall how to solve a system of equations using a fixed-point operator.
Suppose we are given a system of the form
\begin{align*}
  x_0 &= r_0(x_0,\dots,x_{n-1})\,, \\
  &\ \ {}\vdots{} \\
  x_{n-1} &= r_{n-1}(x_0,\dots,x_{n-1})\,,
\end{align*}
where $r_0,\dots,r_{n-1} \in \bbF_n A$.
Inductively defining
\begin{align*}
  s_i(x_0,\dots,x_{i-1}) := (r_i(x_0,\dots,x_i,s_{i+1},\dots,s_{n-1}))^{\omega_i},
\end{align*}
we obtain the new system
\begin{align*}
  x_0 &= s_0\,, \\
  x_1 &= s_1(x_0)\,, \\
      &\ \ {}\vdots{} \\
  x_{n-1} & = s_{n-1}(x_0,\dots,x_{n-2})\,,
\end{align*}
which can now be solved by substitution.

\begin{Prop}\label{Prop: approx_k^m-invariance for trees}
Let\/ $\frakA$~be an algebra for $\cEF_k$. Then
\begin{align*}
  s \approx_k^{(k+3)(\abs{A_0}+1)} t \qtextq{implies} \pi(s) = \pi(t)\,,
  \quad\text{for all regular trees } s,t \in \bbF_0(A_0 \cup A_1)\,.
\end{align*}
\end{Prop}
\begin{proof}
Let $m$~be the number of $\sfL$-classes above $b := \pi(s)$ (including that of~$b$ itself).
We will prove by induction on~$m$ that
\begin{align*}
  s \approx_k^{f(m)} t \qtextq{implies} \pi(t) = b\,,
\end{align*}
where $f(m) := (k+3)(m+1)$.
Set
\begin{align*}
  S &:= \set{ x \in \dom(s) }{ \pi(s|_x) = b }\,, \\
  T &:= \set{ y \in \dom(t) }{ x \approx^{f(m-1)} y \text{ for some } x \in S }\,.
\end{align*}

As $t$~is regular it is the unravelling of some finite graph~$G$.
For each $y \in T$, we will prove that $\pi(t|_y) = b$ by induction on the number of
strongly connected components of~$G$ that are contained in~$T$ and that are reachable from~$y$.
Hence, fix $y \in T$, let $C$~be the strongly connected component of~$G$ containing~$y$,
and choose some $x \in S$ with $x \approx_k^{f(m)-1} y$.
We distinguish two cases.

(a) Let us begin our induction with the case where $C$~is trivial,
i.e., it consists of the single vertex~$y$ without self-loop.
Then
\begin{align*}
  t|_y = a(t_0 +\dots+ t_{n-1} + t'_0 +\dots+ t'_{q-1})
\end{align*}
where $a := t(y)$ and the subtrees~$t_i$ lie outside of~$T$ while the~$t'_i$ contain vertices in~$T$.
Set $d_i := \pi(t_i)$.
By our two inductive hypotheses, we already know that $\pi(t'_i) = b$ and that $b <_\sfL d_i$.
Hence,
\begin{align*}
  \pi(t|_y) = a(d_0 +\dots+ d_{n-1} + q\times b)\,.
\end{align*}
We have to show that this value is equal to~$b$.
Suppose that
\begin{align*}
  s|_x = a(s_0 +\dots+ s_{l-1} + s'_0 +\dots+ s'_{p-1})\,,
\end{align*}
where again the trees~$s_i$ lie outside of~$S$, while the~$s'_i$ contain vertices of~$S$.
Setting $c_i := \pi(s_i)$ it follows that
\begin{align*}
  \pi(s|_x) = a(c_0 +\dots+ c_{l-1} + p\times b)\,.
\end{align*}
Since $x \in S$, we already know that this value is equal to~$b$.
Hence, it remains to show that
\begin{align*}
  a(c_0 +\dots+ c_{l-1} + p\times b) = a(d_0 +\dots+ d_{n-1} + q\times b)\,.
\end{align*}

For $c \in A_0$, let $U_c$~be the set of all vertices $u \succ x$ such that $\pi(s|_u) = c$
and let $V_c$~be the set of vertices $v \succ y$ with $\pi(t|_v) = c$.
As $\leq_\sfL$~is antisymmetric, these sets are convex.
Furthermore, by inductive hypothesis on~$m$, they are also closed under~$\approx_k^{f(m-1)}$.
Since $f(m)-1 = f(m-1)+k+2$,
it therefore follows by Corollary~\ref{Cor: subtrees with value c} that
\begin{align*}
  c_0 +\dots+ c_{l-1} = d_0 +\dots+ d_{n-1}\,.
\end{align*}
If $p = q$, we are done. Hence, we may assume that $p \neq q$.
To conclude the proof, we set
\begin{align*}
  U := \set{ u \in S }{ x \prec u }
  \qtextq{and}
  V := \set{ v \in T }{ y \prec v }\,.
\end{align*}
If $p > 0$, then $x \approx_k^{f(m)-1} y$ and $U \neq \emptyset$ implies $V \neq \emptyset$.
Hence, $q > 0$. In the same way, $q > 0$ implies $p > 0$.
Consequently, we have $p,q > 0$.
We consider several cases.

(i) If $b + b = b$, then
\begin{align*}
  a(d_0 +\dots+ d_{n-1} + q\times b)
  = a(c_0 +\dots+ c_{l-1} + q\times b)
  = a(c_0 +\dots+ c_{l-1} + p\times b)
  = b\,,
\end{align*}
as desired.

(ii)
If $U$~is not a chain, we obtain $b = a'(b,b)$, for some~$a'$,
and Lemma~\ref{Lem: basic properties of leq_L} implies that we are in Case~(i).

(iii)
If $U$~contains an infinite chain, we can use Ramsey's Theorem (or the fact that $s$~is regular),
to obtain a factorisation $b = e^\omega$, which implies that $b + b = b$ by (G3).
Hence, we are in Case~(i) again.

(iv) If $U$~is a finite chain, then so is~$V$, by Lemma~\ref{Lem: number of close ends}.
Hence, $p = 1 = q$ and we are done.

\smallskip
(b)
It remains to consider the case where the component~$C$ is not trivial.
Then we can factorise
\begin{align*}
  t|_y = r(t_0,\dots,t_{n-1},t'_0,\dots,t'_{q-1})\,,
\end{align*}
where $r \in \bbF A$ is the unravelling of~$C$, the subtrees~$t_i$ lie outside of~$T$, while
the subtrees~$t'_i$ contain vertices in~$T$.
Setting $d_i := \pi(t_i)$, it follows by the two inductive hypotheses that
$d_i >_\sfL b$ and $\pi(t'_i) = b$.
Consequently,
\begin{align*}
  \pi(t|_y) = \pi(r)(d_0,\dots,d_{n-1},b,\dots,b)\,.
\end{align*}
Let us simplify the term~$r$.
Introducing one variable~$x_v$, for every vertex $v \in C$, we can write~$r$
as a system of equations
\begin{align*}
  x_v = a_v(x_{u_0} +\dots+ x_{u_{l-1}} + c_0 +\dots+ c_{q-1})\,,
  \quad\text{for } v \in C\,,
\end{align*}
where $u_0,\dots,u_{l-1}$ are the successors of~$v$ that belong to~$C$
and $c_0,\dots,c_{q-1}$ are constants from $\{d_0,\dots,d_{n-1},b\}$
that correspond to successors outside of~$C$.
Solving this system of equations in the way we explained above, we obtain
a finite term~$r_0$ built up from elements of $A_0 \cup A_1$ using
as operations the horizontal product, the vertical product, and the $\omega$-power operation,
such that
\begin{align*}
  \pi(t|_y) = \pi(r_0)(d_0,\dots,d_{n-1},b)\,.
\end{align*}
With the help of the equations (G5)--(G10), we can transform~$r_0$ in several steps
(while preserving its product) until it assumes the form
\begin{align*}
  &\bigl[a_0\cdots a_{j-1}\bigl(x + d_0 +\dots+ d_{n-1} + b\bigr)\bigr]^\omega \\
\prefixtext{or}
  &\bigl[a_0\cdots a_{j-1}\bigl(x + d_0 +\dots+ d_{n-1}\bigr)\bigr]^\omega
\end{align*}
where $a_0,\dots,a_{j-1}$ are the labels of the vertices in~$C$.

We distinguish two cases. First suppose that there is no term with value~$b$
in the above sum. This means that every subtree attached to~$C$ lies entirely outside of
the set~$T$.
Then $x \approx_k^{f(m)-1} y$ implies that we can
factorise~$s|_x$ as
\begin{align*}
  s|_x = r'(s_0,\dots,s_{l-1})
\end{align*}
where
\begin{itemize}
\item $\{\pi(s_0),\dots,\pi(s_{l-1})\} = \{d_0,\dots,d_{n-1}\}$\,,
\item all labels of~$r'$ are among $a_0,\dots,a_{j-1}$,
\item every vertex of~$r'$ has, for every $i < k$, some descendant labelled~$a_i$.
\end{itemize}
As above we can transform~$s|_x$ into
\begin{align*}
  \bigl[a_0\cdots a_{j-1}\bigl(x + c_0 +\dots+ c_{l-1}\bigr)\bigr]^\omega
\end{align*}
where $c_i := \pi(s_i)$.
Since $\{c_0,\dots,c_{l-1}\} = \{d_0,\dots,d_{n-1}\}$ it follows that
\begin{align*}
  \pi(t|_y) &= (a_0\cdots a_{j-1}(x + d_0+\dots+d_{n-1}))^\omega \\
            &= (a_0\cdots a_{j-1}(x + c_0+\dots+c_{l-1}))^\omega = \pi(s|_x) = b\,.
\end{align*}

It thus remains to consider the case where some term has value~$b$.
Using (G7) and (G11) and the fact that $b <_\sfL d_i$,
it then follows that
\begin{align*}
  \pi(t|_y) = \bigl[a_0\cdots a_{j-1}\bigl(x + d_0 +\dots+ d_{n-1} + b\bigr)\bigr]^\omega
            = \bigl[a_0\cdots a_{j-1}(x + b)\bigr]^\omega.
\end{align*}
For every $i < j$, we fix some $z_i \in S$
with label~$a_i$ such that $x \prec z_i$ and some successor of~$z_i$ also belongs to~$S$.
Then
\begin{align*}
  \pi(s|_{z_i}) = a_i(c^i_0 +\dots+ c^i_{l_i-1}+b+\dots+b)\,,
\end{align*}
for some $c^i_0,\dots,c^i_{l_i-1} >_\sfL b$.
Since
\begin{align*}
  b = \pi(s|_{z_i}) = a_i(c^i_0 +\dots+ c^i_{l_i-1}+b+\dots+b)
  \leq_\sfL c^i_0+\dots+c^i_{l_i+1}+b+\dots+b
  \leq_\sfL b
\end{align*}
it follows by asymmetry of~$\leq_\sfL$ that
\begin{align*}
  c^i_0+\dots+c^i_{l_i+1}+b+\dots+b = b
  \qtextq{and}
  a_i(b) = a_i(c^i_0+\dots+c^i_{l_i+1}+b+\dots+b) = b\,.
\end{align*}
Consequently, $a_0\cdots a_{j-1}b = b$, which implies that $a^\pi b = b$
where $a := a_0\cdots a_{j-1}$.
We claim that $b + b = b$. It then follows that
\begin{align*}
  b = a(b) = a(k\times x)(b) = (a(k\times x))^\pi(b)\,,
\end{align*}
which, by (G12)${}_k$, implies that
\begin{align*}
  \pi(t|_y)
   = [a(x+b)]^\omega
   = [a(x + a(k\times x)^\pi(b))]^\omega
   = k\times a(k\times x)^\pi(b)
   = k\times b = b\,,
\end{align*}
as desired.

Hence, it remains to prove our claim that $b + b = b$.
By our assumption on $y$~and~$C$, there is some vertex $u \in C$
that has some successor $v \notin C$ with $v \in T$.
Since $s|_x \approx_k^{f(m)-1} t|_y$ and $f(m) \geq f(m-1)+k+1$,
there are vertices $x \preceq u_0 \prec\dots\prec u_{k-1}$ each of which has some successor $v_i \in S$
with $v_i \npreceq u_{i+1}$.
Consequently, we can write
\begin{align*}
  \pi(s|_x) = a'a''(b,\dots,b)
  \qtextq{and}
  \pi(s|_{u_0}) = a''(b,\dots,b)\,,
\end{align*}
where $a' \in A_1$ and $a'' \in A_k$.
Hence, it follows by (G1)${}_k$ that
\begin{align*}
  b + b = \pi(s|_{u_0}) + b = a''(b,\dots,b) + b = a''(b,\dots,b) = \pi(s|_{u_0}) = b\,.
\end{align*}
\upqed
\end{proof}

\begin{Thm}\label{Thm: cEF algebras bisim invariant}
A regular forest algebra~$\frakA$ is an algebra for $\cEF_k$ if, and only if,
there exists a number $m < \omega$ such that
\begin{align*}
  s \sim_k^m t \qtextq{implies} \pi(s) = \pi(t)\,,
  \quad\text{for all regular forests } s,t \in \bbF(A_0 \cup A_1)\,.
\end{align*}
\end{Thm}
\begin{proof}
$(\Leftarrow)$
In each of the equations (G1)${}_k$--(G12)${}_k$, the two terms on both sides
are $\sim_k^m$-equivalent.

$(\Rightarrow)$
By Proposition~\ref{Prop: approx_k^m-invariance for trees}, there
is some number~$m$ such that
\begin{align*}
  s \approx_k^m t \qtextq{implies} \pi(s) = \pi(t)\,,
  \quad\text{for regular trees } s,t \in \bbF(A_0 \cup A_1)\,.
\end{align*}
Let $s,t \in \bbF(A_0 \cup A_1)$ be regular forests.
We claim that
\begin{align*}
  s \sim_k^{m+k+2} t \qtextq{implies} \pi(s) = \pi(t)\,.
\end{align*}
Suppose that $s = s_0 +\dots+ s_{l-1}$ and $t = t_0 +\dots+ t_{n-1}$,
for trees $s_i$~and~$t_i$, and set $c_i := \pi(s_i)$ and $d_i := \pi(t_i)$.
As in Part~(a) of the proof of Proposition~\ref{Prop: approx_k^m-invariance for trees},
we can use Corollary~\ref{Cor: subtrees with value c} to show that $\pi(s) = \pi(t)$.
\end{proof}

We complete the proof of Theorem~\ref{Thm: cEF-definability} as follows.
\begin{Thm}
A regular language $L \subseteq \bbF_0\Sigma$ is $\cEF_k$-definable if, and only if,
its syntactic algebra~$\frakS(L)$ is an algebra for $\cEF_k$.
\end{Thm}
\begin{proof}
$(\Leftarrow)$ Suppose that $\frakS(L)$~is an algebra for $\cEF_k$.
By Theorem~\ref{Thm: cEF algebras bisim invariant},
every language recognised by~$\frakS(L)$ is invariant under~$\sim_k^m$,
for some~$m$ (when considering regular forests only).
Consequently, the claim follows by Corollary~\ref{Cor: cEF-definable}.

$(\Rightarrow)$ If $L$~is $\cEF_k$-definable, it follows by Corollary~\ref{Cor: cEF-definable}
that $L$~is $\sim_k^m$-invariant, for some~$m$.
Thus $\sim_k^m$~is contained in the syntactic congruence of~$L$, which means that the
syntactic morphism $\eta : \bbF\Sigma \to \frakS(L)$ maps $\sim_k^m$-equivalent
forests to the same value.
Given forests $s,t \in \bbF(S_0 \cup S_1)$ with $s \sim_k^m t$,
we can choose forests $s',t' \in \bbF\Sigma$ with $s' \sim_k^m t'$ and
$s(v) = \eta(s'(v))$ and $t(v) = \eta(t'(v))$.
Thus,
\begin{align*}
  s \sim_k^m t \qtextq{implies} \pi(s) = \eta(s') = \eta(t') = \pi(t)\,.
\end{align*}
By Theorem~\ref{Thm: cEF algebras bisim invariant},
it follows that $\frakS(L)$~is an algebra for~$\cEF_k$.
\end{proof}

{\small
\bibliographystyle{siam}
\bibliography{EF.bib}}

\end{document}